\documentclass[aps,prl,twocolumn,superscriptaddress,showpacs,floatfix]{revtex4-2}

\usepackage{amsmath,amssymb,amsfonts,float,graphics,epsfig,epstopdf,color,verbatim,tabularx,bm,multirow,appendix}
\usepackage[utf8]{inputenc}
\usepackage[T1]{fontenc}
\usepackage{xcolor}
\usepackage{dsfont}
\usepackage{textcomp}
\usepackage{yfonts}
\usepackage{footnote}
\usepackage{bm}
\usepackage{subfigure}
\usepackage{mathrsfs}
\usepackage{graphicx}
\usepackage{verbatim}
\usepackage{hyperref}
\usepackage{multirow}
\usepackage{braket}
\usepackage[normalem]{ulem}
\usepackage{tikz}
\usetikzlibrary{calc}
\usetikzlibrary{shapes.multipart}

\renewcommand{\vec}[1]{{\mathbf #1}}

\newcommand{\comments}[1]{}

\newcommand{\stkout}[1]{\ifmmode\text{\sout{\ensuremath{#1}}}\else\sout{#1}\fi}

\makeatletter
\def\l@subsubsection#1#2{}
\makeatother

\begin{document}

\title{Dynamical properties of quantum many-body systems with long-range interactions}

\author{Menghan Song}
\affiliation{Department of Physics and HKU-UCAS Joint Institute of Theoretical and Computational Physics, The University of Hong Kong, Pokfulam Road, Hong Kong SAR, China}

\author{Jiarui Zhao}
\email{jrzhao@connect.hku.hk}
\affiliation{Department of Physics and HKU-UCAS Joint Institute of Theoretical and Computational Physics, The University of Hong Kong, Pokfulam Road, Hong Kong SAR, China}

\author{Chengkang Zhou}
\email{zhouchk@connect.hku.hk}
\affiliation{Department of Physics and HKU-UCAS Joint Institute of Theoretical and Computational Physics, The University of Hong Kong, Pokfulam Road, Hong Kong SAR, China}


\author{Zi Yang Meng}
\email{zymeng@hku.hk}
\affiliation{Department of Physics and HKU-UCAS Joint Institute of Theoretical and Computational Physics, The University of Hong Kong, Pokfulam Road, Hong Kong SAR, China}

\begin{abstract}
Employing large-scale quantum Monte Carlo simulations, we systematically compute the energy spectra of the 2D spin-1/2 Heisenberg model with long-range interactions. With the $1/r^{\alpha}$ ferromagnetic and staggered antiferromagnetic interactions, we find the explicit range in $\alpha$ for {\color{black} the short-range Goldstone-type (gapless), anomalous Goldstone-type (gapless) and Higgs-type (gapped) spectra. Accompanied by the spin wave analysis, our numerical results vividly reveal how the long-range interactions alter the usual linear and quadratic magnon dispersions in 2D quantum mangets and give rise to anomalous dynamical exponents. Moreover, we find explicit case where the gapped excitation exists even when the Hamiltonian is extensive. This work provides the first set of unbiased dynamical data} of long-range quantum many-body systems and suggests that many universally accepted low-energy customs for short-range systems need to be substantially modified for long-range ones which are of immediate relevance to the ongoing experimental efforts from quantum simulators to 2D quantum moir\'e materials.
\end{abstract}

\date{\today}
\maketitle

\noindent{\textcolor{blue}{\it Introduction.}---}
Quantum many-body systems with long-range (LR) interactions exhibit different and exotic properties compared with their short-range cousins, as the LR nature of the interaction differentiates them from many universally accepted long-wavelength and low-energy customs governing the short-range ones over the years. For example, the well-known Hohenberg-Mermin-Wagner theorem~\cite{hohenbergExistence1967,merinAbsence1966} that forbids spontaneous symmetry-breaking of continuous symmetry at finite temperature in low dimensions can be easily circumvented and LR interactions can generate interesting finite temperature transitions~\cite{WeberDissipation2022,WangOn2022,wernerQuantum2005,LiaoCaution2022,halperin2019,LiLong2021,PeterAnomalous2012} and new critical phenonema~\cite{FisherCritical1972,SakRecursion1973,DefenuLong2021,KoziolQuantum2021,FeyQuantum2019,AdelhardtContinuously2022}. The bedrock in the research of highly entangled quantum matter -- the area law scaling of the entanglement entropy --  can also be bypassed in LR systems, and the consequent new scaling behavior points towards new guiding principle of quantum entanglement that awaits to be worked out~\cite{KoffelEntanglement2012,LiLong2021,jiangFermion2022,zhaoScaling2022,wangScaling2022,liuFermion2022,liaoTeaching2023}.


Moreover, recently the field of LR quantum-many systems becomes even more active due to their fast experimental realizations, such as the Rydberg atom arrays with long-range van der Waals or dipolar couplings where {\color{black}topological ordered phases, emergent glassy behavior and quantum criticality have been suggested and realized~\cite{samajdar2021quantum,yan2022triangular,Semeghini21,Roushan21,yanEmergent2023}, magic angle twisted bilayer Graphene (TBG) and 2D quantum moir\'e materials in which flat-band topology and long-range Coulomb interaction give rise to a plethora of correlated phases beyond semi-classical or band-theory description~\cite{tramblyLocalization2010,rafiMoire2011,lopesContinuum2012,caoUnconventional2018,caoCorrelated2018,xieSpectroscopic2019,luSuperconductors2019,kerelskyMaximized2019,liaoValence2019,yankowitzTuning2019,yankowitzTuning2019,tomarkenElectronic2019,caoStrange2020,shenCorrelated2020,chatterjeeSkyrmion2022,khalafSoftmodes2020,xieNature2020,pierceUnconventional2021,liaoCorrelated2021,rozenEntropic2021,zondinerCascade2020,saitoIsospin2021,parkFlavour2021,kwanExciton2021,liaoCorrelation2021,liuTheories2021,schindlerTrion2022,songMagic2022,linSpin2022,huangObservation2022,zhangCorrelated2022,herzogReentrant2022,andreiGraphene2020,stepanovCompeting2021,panThermodynamic2022,zhangMomentum2021,hofmannFermionic2022,panDynamical2021,zhangFermion2022,zhangSuperconductivity2022,zhangQuantum2022,chenRealization2021,linExciton2022,huangIntrinsic2022}, as well as the quantum gases coupled to optical cavities~\cite{RitschCold2013} and programmable quantum simulators~\cite{verresenPrediction2021,samajdarEmergent2022,yanFully2022,ranFully2022,wangFractionalized2021,yanEmergent2023}.}

\begin{figure}[htp!]
	\includegraphics[width=0.95\columnwidth]{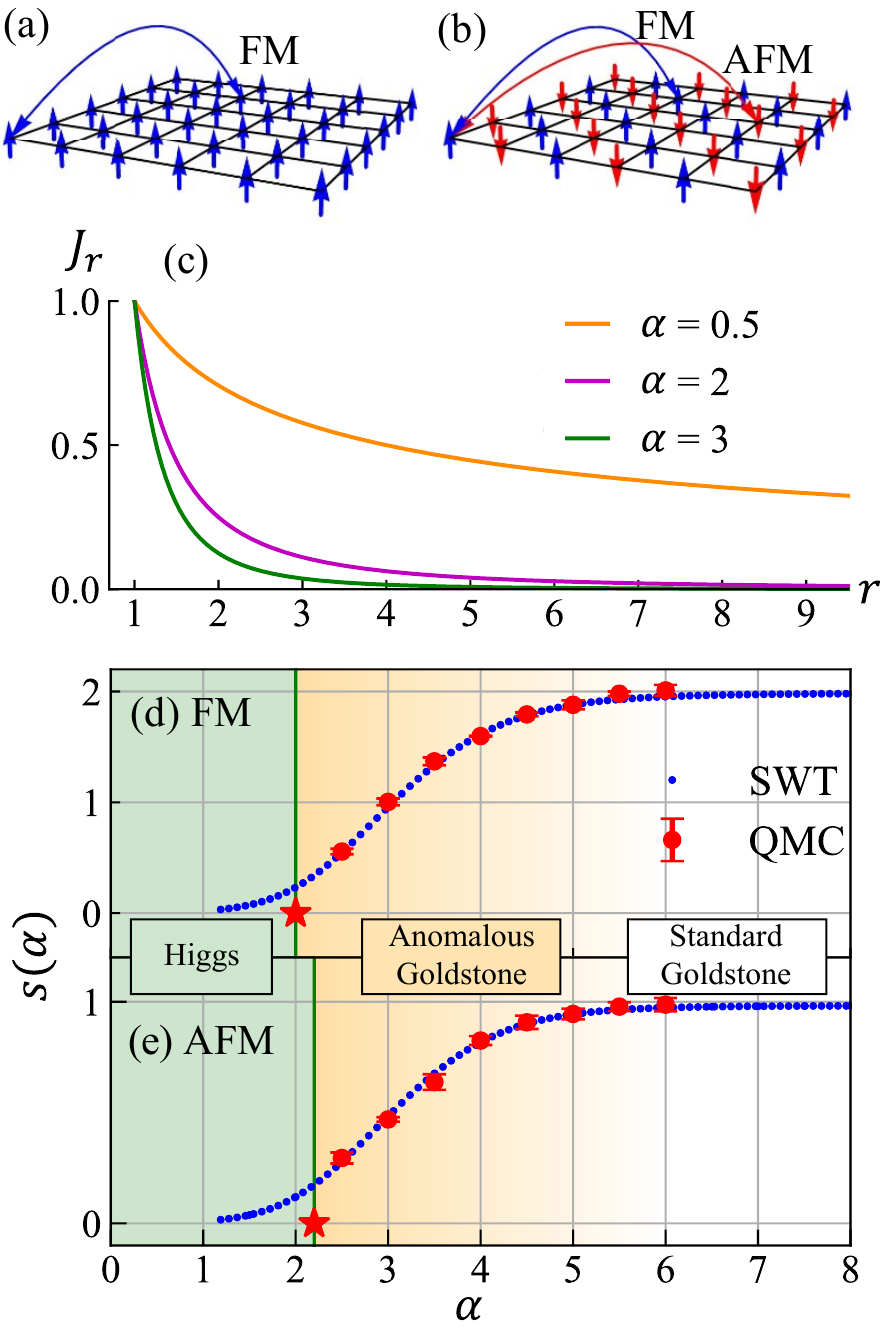}
	\caption{\textbf{2D LR Heisenberg models with Higgs, anomalous and short-range Goldstone spectra.} The schematic plots of 2D Heisenberg model with LR ferromagnetic interaction (a) and staggered antiferromagnetic interaction (b). The power-law decay of $J(r) \sim 1/r^{\alpha}$ for three different $\alpha$ is shown in (c). (d) and (e) show the power $s(\alpha)$ of low-energy spectra $\omega \sim |\mathbf{q}|^{s(\alpha)}$ obtained from QMC and SWT versus $\alpha$ for both the ferromagnetic and antiferromagnetic cases. The green-shaded area represents the Higgs regime where the spectra are gapped, the yellow shaded area represents the anomalous Goldstone regime where the dispersion powers change with $\alpha$, and the white area is the standard {\color{black}short-range Goldstone regime where $s=1$ for antiferrmagnetic and $s=2$ for ferromagnetic cases.} The red dots are fitting results from QMC ($L=64$) and the red stars denote QMC boundaries at $\alpha=2$ (for ferromagnetic) and $\alpha=2.2$ (for antiferromagnetic) which speparate the Higgs and Goldstone regimes. The blue dashed lines are fitting results from SWT, with a cut-off of longest coupling distance $r_{\mathrm{max}}=1000$.}
	\label{fig:fig1}
\end{figure}

Despite such fast developments,  theoretical and numerical investigations on the dynamical properties of the LR quantum many-body systems are however still lacking. This is mainly due to the fact that dynamical properties, such as spectral functions~\cite{shaoNearly2017,maDynamical2018,sunDynamical2018,wangDynamics2019,yanTopological2021,zhouAmplitude2021,panDynamical2021,zhouEvolution2022,yanWormhole2021,songReversing2022}, are usually difficult to compute without approximation in analytic theory and numerical simulations, even for the short-ranged systems. And therefore by now there only exist few perturbative works such as Refs.~\cite{DiesselGeneralized2022,chiocchettaCavity2021,EddySpin2004,DefenuMetastability2021,BirnkammerCharacterzing2022}, which are mainly valid at various mean-field limits where the fluctuations are suppressed, and previous algorithmic developments in non-perturbative numerical approaches for LR system are mainly focused on classical and 1D systems~\cite{FukuiOrder2009,HoritaUpper2017}. However, in the aforementioned experiments of 2D quantum LR systems, it is actually the dynamical information that can be easily detected by means of neutron scattering, nuclear magnetic resonance, scanning tunneling spectroscopy, nonlinear and non-equilibrium transport and optical probes, etc.

To overcome the dilemma between the fast experimental developments and the slow theoretical reality in LR quantum many-body systems, the need to develop and carry out unbiased approaches such as large-scale quantum Monte Carlo (QMC) simulations, to systematically investigate the dynamical properties therein is obvious. And only in this way, can one fully reveal the interplay between the LR interaction and quantum topology and fluctuations to explain the fascinating experimental outcomes and predict new ones.

\begin{figure*}[htp!]
	\centering
	\includegraphics[width=1\textwidth]{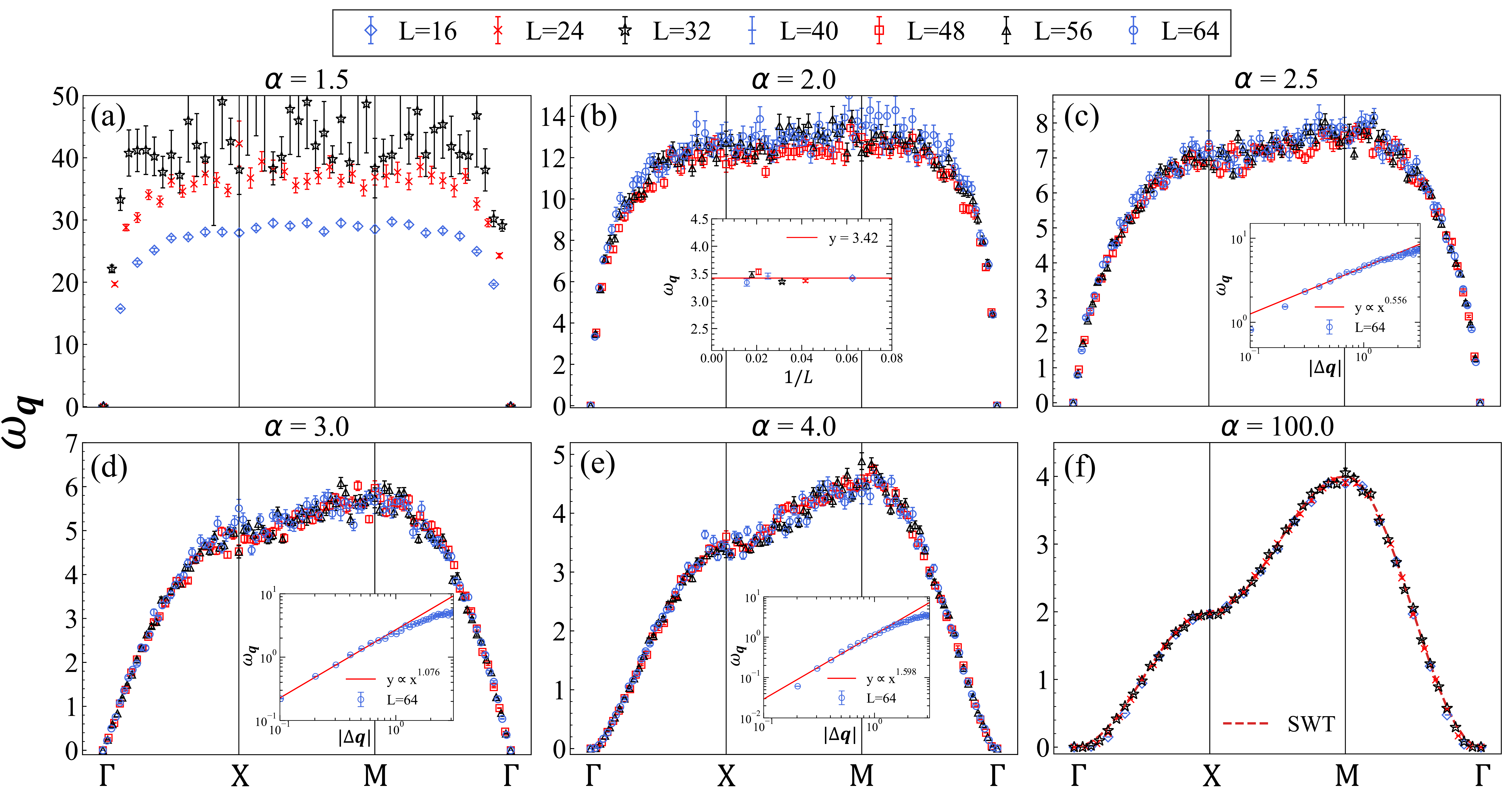}
	\caption{\textbf{Dynamical properties of 2D LR ferromagnetic Heisenberg model.} Dispersion relations along the path ($\Gamma\to X \to M \to \Gamma$) with panels (a)-(f) for different decay exponents $\alpha$. Results of various sizes $L$ are plotted together in each panel and share the same legend on the top. Inset of panel (b) indicates that at $\alpha=2$ the first excitation gaps near $\Gamma$ for various sizes converge to a finite value and the system has a gaped spectrum, i.e., inside the Higgs-regime. Insets of (c), (d) and (e) show the fitting of power-law dispersions $\omega_{\vec{q}} \sim |\mathbf{q}|^{s(\alpha)}$ near ${\Gamma}$ (with $|\Delta \mathbf{q}|$ denotes the relative momentum away from $\Gamma$) in range $2 < \alpha \le 4$. Red dashed line in (f) is the SWT dispersion for 2D nearest neighbor FM Heisenberg model with $\omega _\vec{q} =\left |J  \right |zS(1-\gamma_\vec{q})$ where $S=1/2$, the coordination number $z=4$, and $\gamma_\vec{q} =\frac{1}{z}\sum_{\delta} e^{i\vec{q}{\delta}}$. }
	\label{fig:fig2}
\end{figure*}

This is the focus of our paper. Here we develop and employ the stochastic series expansion (SSE) QMC~\cite{SandvikQuantum1991,SandvikStochastic1999,yanEmergent2023,yanWormhole2021} simulation for the LR quantum many-body systems, to compute the energy spectra of the 2D spin-1/2 Heisenberg model with $1/r^{\alpha}$ interaction where $\alpha$ is the decay exponent, as shown in Fig.~\ref{fig:fig1}. With the interaction types of ferromagnetic (see Fig.~\ref{fig:fig1} (a)) and antiferromagnetic (staggered without introducing frustration, Fig.~\ref{fig:fig1} (b)), we find the explicit range in $\alpha$ for {\color{black}the short-range Goldstone-type (gapless), anomalous Goldstone-type (gapless but with varying dynamical exponent) and Higgs-type (where the spectra are gapped) spectra.} As shown in Fig.~\ref{fig:fig1} (d) and (e), accompanied by spin-wave theory (SWT) analysis~\cite{TothLinear2015,DefenuLong2021,vannote1980,HamerThird-order1992,ZhengSeries2005,SyromyatnikovSpectrum2010}, our results reveal how the long-range interactions induce a mass to the Goldstone mode via the generalized Higgs mechanism~\cite{DiesselGeneralized2022}. {\color{black}Moreover, different from the conventional wisdom for systems with long-range interactions~\cite{VanKac1963,MickaelClusting1995,DefenuMetastability2021,MichaeUniversal2013}, we find explicit case -- the staggered antiferromagnetic model -- where the gapped excitation exists even when the Hamiltonian is extensive. Therefore our work provides the first set of unbiased dynamical data of LR quantum many-body systems where universally accepted low-energy physics} are substantially modified. Implications of ongoing experiments in quantum simulators and 2D quantum moir\'e materials are discussed.

\noindent{\textcolor{blue}{\it Model and Method.}---}
We consider the 2D spin-1/2 LR Heisenberg model with power-law decaying couplings on the square lattice. The Hamiltonians for the ferromagnetic and the antiferromagnetic cases (with staggered interaction to avoid the sign problem~\cite{LohSign1990,panSign2022}) are given by
\begin{eqnarray}
	H_{FM}&=&-J\sum_{i\neq j}\frac{1}{|\vec{r_{i}-\vec{r_{j}}|^{\alpha}}} \vec{S_{i}}\cdot
	\vec{S_{j}},
	\label{ferroH}\\
	H_{AFM}&=&J\sum_{i\neq j}\frac{(-1)^{|x_i+y_i-x_j-y_j+1|}}{|\vec{r_{i}-\vec{r_{j}}|^{\alpha}}} \vec{S_{i}}\cdot
	\vec{S_{j}}.
	\label{antiferroH}
\end{eqnarray}
The schematic spin configurations and the decaying LR interactions of $J(r)=1/r^{\alpha}$ for {\color{black}both cases are shown in Fig.~\ref{fig:fig1}(a-c). Here we set $J=1$ and simulate the system sizes upto $L=64$, the inverse temperature $\beta=L/2$ and the decay exponent $\alpha$ from 1.5 to 100, with the focus on $\alpha \le 6$. We note that to probe the ground-state properties in finite space-time size QMC simulations, one usually scale $\beta\sim L^{z}$ where $z$ is the dynamical exponents. In our problem, $z$ is the largest at the short-ranged cases (for example, $z=1$ for short-range antiferromagnetic case) and our results are well converged. Since the long-range cases will only have smaller $z$, $\beta=L/2$ is more than sufficient to extract the ground state properties therein. Detailed implementation and finite size analysis of the obtained dispersions in QMC are shown in Sec. II of Supplementary Materials (SM)~\cite{suppl}.}

As discussed in Refs.~\cite{BrunoAbsence2001,DiesselGeneralized2022}, for $H_{FM}$, the SWT analysis accompanied by a continuum approximation concludes that for $\alpha > d+2$ (denoted as standard Goldstone regime) where $d$ is the spatial dimension, the low-momentum dispersion of the LR model reduces to the short-range case with $\omega \sim |\mathbf{q}|^2$, and for $d < \alpha < d+2$ (denoted as anomalous Goldstone regime) the dispersion is $\omega \sim |\mathbf{q}|^{\alpha-d}$. For $\alpha \le d$ (denoted as Higgs regime) the system becomes gapped because of the generalized Higgs mechanism~\cite{DiesselGeneralized2022}. As will be shown below, our QMC results are {\color{black}consistent with this picture as we reveal three different regimes via fitting $\omega \sim |\mathbf{q}|^{s(\alpha)}$ and finite size analysis. When $\alpha$ is large, the system is in the standard Goldstone region with dispersion $\omega \sim |\mathbf{q}|^2$. As $\alpha$ decreases, LR interaction brings the system into the anomalous Goldstone regime, and we find exactly the same Higgs regime as in Ref.~\cite{DiesselGeneralized2022} which is $\alpha \le 2$ (see Fig.~\ref{fig:fig1} (d))}

\begin{figure*}[htp!]
	\centering
	\includegraphics[width=\textwidth]{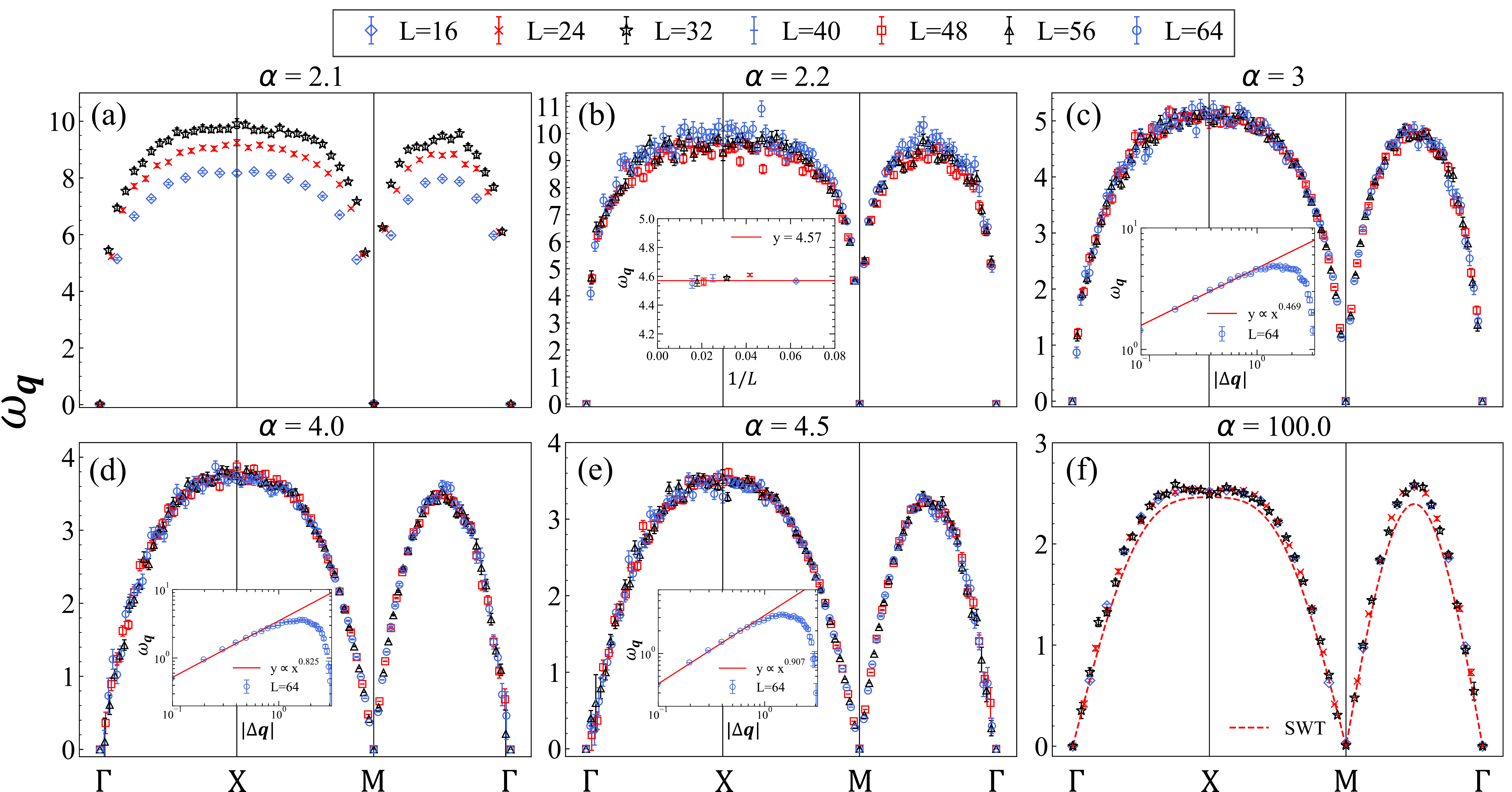}
	\caption{\textbf{Dynamical properties of 2D LR (staggered) antiferromagnetic Heisenberg model.} Dispersion relations along the path ($\Gamma\to X \to M \to \Gamma$) with panels (a)-(f) for different decay exponents $\alpha$. Results of various sizes $L$ are plotted together in each panel and share the same legend on the top. Inset of panel (b) indicates that the first excitation gaps near $M=(\pi,\pi)$ for various sizes converge to a finite value and thus the system is inside the Higgs regime at $\alpha\le 2.2$. Insets of (c), (d) and (e) show the fitting of power-law dispersion as $\omega \sim |\vec{q}|^{s(\alpha)}$ near $M$ (with $|\Delta \mathbf{q}|$ denotes the relative momentum away from $M$) in range $2.2 < \alpha < 4.5$. Dashed red line in (f) is the nearest-neighbor SWT dispersion $\omega _\vec{q} =\left |J  \right |zS\sqrt{(1-\gamma_\vec{q})^2}$ with an additional coefficient $\sim$1.158 to approximate the second order spin wave effects~\cite{Weihong2005,CJHamer1992,shaoNearly2017,SyromyatnikovSpectrum2010}.}
	\label{fig:fig3}
\end{figure*}

As for the antiferromagnetic case, it is worth noting that Ref.~\cite{DiesselGeneralized2022} predicts the Higgs regime occurs at $\alpha\le d-2$ for the Hamiltonian  $H=J\sum_{i\neq j}\frac{1}{|\vec{r_{i}-\vec{r_{j}}|^{\alpha}}} \vec{S_{i}}\cdot
\vec{S_{j}}$. Therefore for $d=2$ there will be no finite $\alpha$ values with gapped spectra, and the anomalous and standard Goldstone regimes are $d-2< \alpha < d$ and $\alpha > d$ respectively. However, we consider a sign-problem-free  Hamiltonian of Eq.~\eqref{antiferroH} which does not host frustrations and {\color{black}we get different boundaries of the three regimes. The system returns to the standard Goldstone mode when $\alpha$ is large enough, and our QMC results show the anomalous Goldstone regime is $\alpha >2.2$ and the Higgs regime is $\alpha \le 2.2$ (see Fig.~\ref{fig:fig1} (e)). 
	
It is interesting to see both $H_{FM}$ and $H_{AFM}$ are superextensive when $\alpha\le2$ and it can be seen from Fig.~\ref{fig:fig2} (a) that the gap diverges with system sizes at $\alpha=1.5$ in the ferromagnetic case. Between $2<\alpha<2.2$ for $H_{AFM}$, the Hamiltonian is extensive and yet the spectra are gapped (as shown in Fig.~\ref{fig:fig3} (a) and (b)). Therefore, our results go beyond the conventional wisdom for long-range systems~\cite{VanKac1963,MickaelClusting1995,DefenuMetastability2021,MichaeUniversal2013}, and suggesed the disentanglement of the gapped spectrum and the superextensiveness of the Hamiltonian. We further considered the Kac construction which couples a normalization factor $(N-1)/\sum_{i\ne j}\frac{1}{\left | r_i-r_j \right | ^{\alpha}}$ to Eqs.~\eqref{ferroH} and perform SWT analysis on the normalized Hamiltonian and confirm the gapped spectra with the same boundaries. SWT data are presented in Sec.I in SM~\cite{suppl}}.   


\noindent{\textcolor{blue}{\it Results.}---} Fig.~\ref{fig:fig2} shows the obtained QMC spectra along the high symmetry path $\Gamma(0,0)\to X(\pi,0) \to M(\pi,\pi) \to \Gamma(0,0)$ for the ferromagnetic case. At $\alpha=100$ (panel (f)), the system reduces to the short-ranged case with only nearest-neighbor couplings~\cite{FisherCritical1972,SakRecursion1973,DefenuLong2021}, and our QMC-obtained spectra matches well with the SWT spectra. Both of them show a $\omega_{\vec{q}} \sim |\vec{q}|^2$ dispersion close to $\Gamma$, {\color{black}and they match well along the whole path.} As $\alpha$ gets smaller, as shown in panels (c), (d) and (e), we find the dispersion enters the anomalous Goldstone region~\cite{DiesselGeneralized2022}, i.e., the dispersion close to $\Gamma$ deviates from a quadratic one. We use $\omega_{\vec{q}} \sim |\mathbf{q}|^{s(\alpha)}$ to fit the dispersion close to $\Gamma$ and find the power $s(\alpha)$ gradually decreases as $\alpha$ gets smaller. Insets of these three panels demonstrate the power law fitting of $s(\alpha)$ using $L=64$ QMC data. We find, at $\alpha=3$ (panel (d)),  $s=1.076$ which agrees well with the relation of $s(\alpha)=\alpha-2$ suggested in Ref.~\cite{DiesselGeneralized2022}. However, for $\alpha=4$ (panel (e)) and $\alpha=2.5$ (panel c)) our results show apparent derivations from $s(\alpha)=\alpha-2$. Fig.~\ref{fig:fig1}(d) collects the fitted power $s(\alpha)$ by QMC (red dots) at various $\alpha$ and we observe a satisfactory match with our SWT results. At $\alpha=2$, we find $\omega_{\vec{q}}$ near $\Gamma$, i.e., $\vec{q}=(\frac{2\pi}{L},0)$ for different system sizes converge to a large and finite value of $\omega \approx 3.42$ as indicated in the inset of Fig.~\ref{fig:fig2} (b). This phenomenon is fundamentally different from a gapless excitation in which the finite size gap $\omega_{({2\pi}/{L},0)}$ converges to zero as $L\to \infty$ and results in a continuous spectra. Our result reveals that at $\alpha=2$ the system enters the Higgs regime where the Goldstone mode acquires mass due to the LR interation and the excitation spectrum becomes gapped. {\color{black}For $\alpha<2$ ($\alpha=1.5$ in panel (a)) we find the gaps begin to diverge with the system size $L$ due to the aforementioned super-extensive Hamiltonian.} Therefore, we conclude that $\alpha=2$ is the separation power between the Higgs-type and Goldstone-type spectra in $H_{FM}$ from our QMC results. 

Fig.~\ref{fig:fig3} illustrates the QMC dispersion relation for $H_{AFM}$ along the high symmetry path. Similarly, in panel (f), we benchmark the spectrum at $\alpha=100$ with SWT result for the short-range antiferromagnetic Hamiltonian (with an extra coefficient $\sim$1.158 multiplied to approximate the second order spin wave effects~\cite{Weihong2005,CJHamer1992,shaoNearly2017,SyromyatnikovSpectrum2010}) and find QMC results agree well with SWT dispersion close to $M$ with $\omega_{\vec{q}} \sim |\vec{q}|$. As $\alpha$ decreases, the system also enters the anomalous Goldstone region with $\omega_{\vec{q}} \sim |\vec{q}|^{s(\alpha)}$ and $0< s(\alpha)<1$ close to $M$. Fitted powers via QMC at various $\alpha$ are displayed in Fig.~\ref{fig:fig1}(e) and agree well with the SWT results. In Fig.~\ref{fig:fig3} (b) at $\alpha=2.2$, $\omega_{\vec{q}}$ close to $M$ converges to a large and finite value of $\omega=4.57$. {\color{black}This means $H_{AFM}$ is in the Higgs-regime with gapped spectra when $\alpha\le 2.2$. And in fact, panel (a) with $\alpha=2.1$ clearly forecasts a non-vanishing gap close to $M$ at thermaldynamic limit. Interestingly, performing similar analysis as done in Ref.~\cite{DiesselGeneralized2022}, one would obtain the same Higgs boundary as in the ferromagnetic case, i.e. $\alpha=2$ which deviates from the boundary at $\alpha=2.2$ obtained from unbiased QMC simulations. The deviation seems to suggest that antiferromagnetic quantum fluctuation plays a non-negligible role in this gap generating process and makes it different from the ferromagnetic case. Therefore, it is of interest to conduct further theoretical analysis for Eq.~\eqref{antiferroH} to understand this discrepancy and reveal the subtle working of the disentanglement of the gapped spectrum and the super-extensiveness of the Hamiltonian.}


\noindent{\textcolor{blue}{\it Discussions.}---}
With the unbiased large-scale QMC simulations and SWT analysis, we systematically investigate the dynamical properties of 2D spin-1/2 Heisenberg model with LR interactions. We find in contrast to the well accepted low-energy customs such as Hohenberg-Mermin-Wagner theorem and gapless Goldstone mode, the LR quantum many-body systems offer richer tunability and exhibit new phenomena. As the interaction exponent $\alpha$ varies, the Goldstone modes can be strongly modified, in that they can be either distorted (in the anomalous Goldstone regime), or even be gapped via a generalized Higgs mechanism. {\color{black}We also find explicit case where the gapped excitation exists even when the Hamiltonian is extensive. }

These dynamical properties have immediate relevance to the ongoing experiments with ultracold atom arrays and quantum moir\'e materials. For example, the long-range  Coulomb interaction in quantum moir\'e systems can be easily tuned by varying dielectric environment, electrostatic gating and twisting angles, and in this way observed thermodynamical and dynamical properties (such as switching between gapped and gapless spectra)~\cite{linExciton2022,panThermodynamic2022,zhangSuperconductivity2022,zhangQuantum2022,stepanovCompeting2021,kerelskyMaximized2019} can be identified with different LR interaction types and regimes, when compared unbiased results such as ours. Similar tunability can also be realised in dressed Rydberg atom arrays whose interaction can be modified~\cite{Rydberg_dress}, one can then compare different responses from experiments with our results to identify the LR interaction and the novel phases. 


{\it{Acknowledgment.-}} We thank Zheng Yan, Tianyu Wu, Ting-Tung Wang, Yuan Da Liao, Qi Yang, Meng Cheng and Fakher Assaad for valuable discussions on related topics. We acknowledge the support from the Research Grants Council of Hong Kong SAR of China (Project Nos. 17301420, 17301721, AoE/P-701/20, 17309822, HKU C7037-22G), the ANR/RGC Joint Research Scheme sponsored by Research Grants Council of Hong Kong SAR of China and French National Research Agency(Project No. A\_HKU703/22), the K. C. Wong Education Foundation (Grant No.~GJTD-2020-01) and the Seed Funding “Quantum-Inspired explainable-AI” at the HKU-TCL Joint Research Centre for Artificial Intelligence. We thank the HPC2021 system under the Information Technology Services and the Blackbody HPC system at the Department of Physics, the University of Hong Kong for their technical support and generous allocation of CPU time. The authors also acknowledge the National Supercomputer Centers in Guangzhou, the Beijng PARATERA Tech CO.,Ltd.(\url{https://www.paratera.com/}) for providing HPC resources that have contributed to the research results reported within this paper.

\bibliography{LR_spectrum}
\appendix

\newpage
\newpage
	
\section{Supplementary Materials}
In this supplementary material, we present the linear spin wave analysis for the LR FM and staggered AFM Heisenberg model, in which the dispersion relation of the low-lying magnetic excitations at different decaying power $\alpha$ are extracted. From here, we make comparison with the dispersion obtained from the QMC simulations in the main text. \textcolor{black}{Spin wave analysis with Kac construction is presented. Moreover, we outline the QMC procedure and provide detailed data on the fitting of the excitation gaps from the dynamical correlation functions obtained in QMC simulations.}

\subsection{Linear spin wave analysis}
\begin{figure*}[htpt]
	\includegraphics[width=1.85\columnwidth]{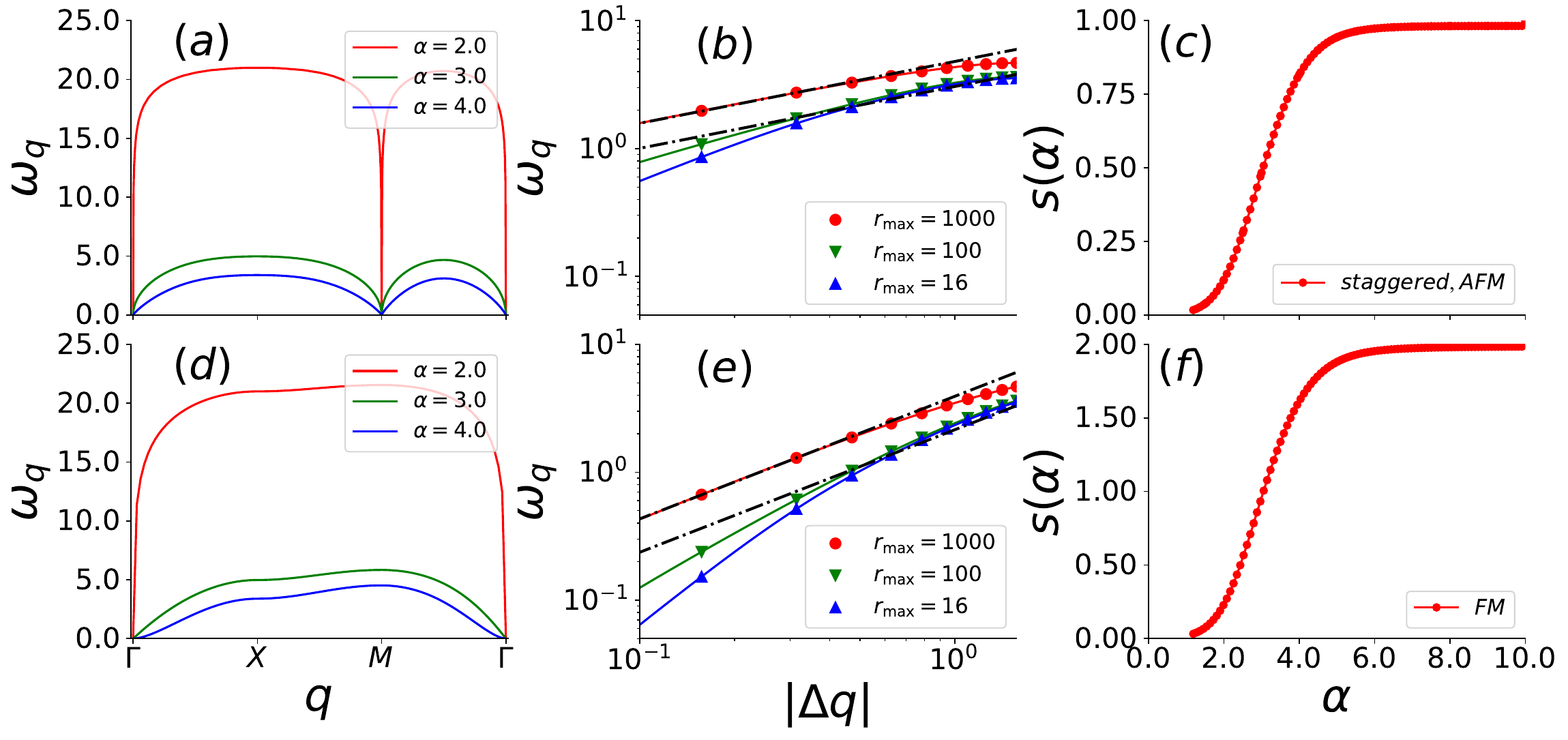}
	\caption{The linear SWT results. Panels (a), (b), and (c) are the linear spin wave result of the staggered antiferromagnetic case while (d), (e), and (f) are the ferromagnetic case. Panel (a) and (d) are the dispersion relation plotted along the momentum path $\Gamma\rightarrow X\rightarrow M\rightarrow \Gamma$ with $r_{\mathrm{max}}=1000$. (b) is the spin wave dispersion relation near the $M$ point for the staggered antiferromagnetic lattice with $\alpha=3.5$ and $r_{\mathrm{max}}$ ranges from $16$ to $1000$ while (e) for $\Gamma$ point in the ferromagnetic case. And two black dashed lines here refer to the relation $\omega_{\vec{q}}\propto|\Delta\vec{q}|^{0.48}$ in (b) and  $\omega_{\vec{q}}\propto|\Delta\vec{q}|^{0.96}$ in (e). (c) and (f) describe the relation between the power of the dispersion $s(\alpha)$ as a function of decay exponent $\alpha$, obtained from fittings with $r_{\mathrm{max}}=1000$.}
	\label{fig:axfig1}
\end{figure*}

We applied the linear spin wave theory (SWT) to analyze the dispersion of the low energy excitation in the LR spin-1/2 Heisenberg model with power law decaying couplings in both ferromagnetic and staggered antiferromagnetic cases~\cite{TothLinear2015,DefenuLong2021,vannote1980,HamerThird-order1992,ZhengSeries2005,SyromyatnikovSpectrum2010}. Taking the staggered antiferromagnetic cases as an example, it calls for the definition of two sublattices, $A$ and $B$. The spin on each sublattice is pointing in the same direction. Then, we rewrite the spin operators by $S^{+}=S^{x}+iS^{y}$ and $S^{-}=S^{x}-iS^{y}$ and apply the Holstein-Primakoff transformation up to order $S$ that for sublattice $A$
\begin{equation}
	\begin{aligned}
		S^{z}_{i}&=S-a^{\dagger}_{i}a^{}_{i},\\
		S^{+}_{i}&=\sqrt{2S}a^{}_{i},\\
		S^{-}_{i}&=\sqrt{2S}a^{\dagger}_{i},\\
	\end{aligned}
	\label{appLSW:HPA}
\end{equation}and for sublattice $B$
\begin{equation}
	\begin{aligned}
		S^{z}_{i}&=b^{\dagger}_{i}b^{}_{i}-S,\\
		S^{+}_{i}&=\sqrt{2S}b^{\dagger}_{i},\\
		S^{-}_{i}&=\sqrt{2S}b^{}_{i}.\\
	\end{aligned}
	\label{appLSW:HPB}
\end{equation}
 Here we take $S=1/2$ and the Hamiltonian in the momentum space is given by
\begin{equation}
	\begin{aligned}
		H_{sw}&=\sum_{\mathbf{q}}\gamma^{\dagger}(\mathbf{q})H_{\mathbf{q}}\gamma(\mathbf{q}),\\
		H_{\mathbf{q}}&=\begin{bmatrix}
		J^{d}_{0}+J^{s}_{0}-J^{s}_{\mathbf{q}}& J^{d}_{\mathbf{q}}\\
		J^{d}_{\mathbf{q}}& J^{d}_{0}+J^{s}_{0}-J^{s}_{\mathbf{q}}
		\end{bmatrix},
	\end{aligned}
	\label{appLSW:Ham}
\end{equation}
in which $\gamma^{\dagger}(\mathbf{q})=(a^{\dagger}_{\mathbf{q}},b^{}_{\mathbf{q}})$ and $a^{\dagger}_{\mathbf{q}}$ is the Fourier transformed that $a^{\dagger}_{\mathbf{q}}=N^{1/2}\sum_{\mathbf{q}}a^{\dagger}_{i}e^{-i\mathbf{q}\mathbf{r}}$.
And, $J^{s}_{\mathbf{q}}=\sum_{\mathbf{r}^{s}\in \mathrm{same}}e^{-i\mathbf{q}\mathbf{r}^{s}}J^{s}_{\mathbf{r}}$ refers to the coupling between the spins belong to the same sublattice, and $J^{d}_{\mathbf{q}}=\sum_{\mathbf{r}^{d}\in \mathrm{diff}}e^{-i\mathbf{q}\mathbf{r}^{d}}J^{d}_{\mathbf{r}}$ to that of the different sublattices. $J^{d(s)}_{\mathbf{r}}=1/|\Delta r|^{\alpha}$ is the coupling strength. Finally, the single magnon dispersion relation of the LR Heisenberg model with staggered antiferromagnetic power-law decaying couplings is given by
\begin{equation}
		\omega^{\mathrm{AFM}}_{\mathbf{q}}=
		\sqrt{(J^{d}_{0}+J^{s}_{0}-J^{s}_{\mathbf{q}}+J^{d}_{\mathbf{q}})(J^{d}_{0}+J^{s}_{0}-J^{s}_{\mathbf{q}}-J^{d}_{\mathbf{q}})}.
	\label{appLSW:omega}
\end{equation}

Similarity, in the ferromagnetic case, the dispersion relation of single magnon can be read as $\omega^{\mathrm{FM}}_{\mathbf{q}}=|J_{0}-J_{\mathbf{q}}|$ with $J_{\mathbf{q}}=\sum_{\mathbf{r}}e^{-i\mathbf{q}\mathbf{r}}J_{\mathbf{r}}$.

To capture the dependence of the dispersion relation to $\alpha$ in Eqs.~\eqref{ferroH} and \eqref{antiferroH} in the main text, we numerically calculate the linear SWT results by applying a cut-off of longest range coupling as $r_{\mathrm{max}}$, meaning that we only consider the coupling between the sites $(r_{x},r_{y})$ and $(r_{x}+\Delta r_{x},r_{y}+\Delta r_{y})$ with $\Delta r_{x}$ and $\Delta r_{y}$ ranging from $-r_{\mathrm{max}}/2$ to  $r_{\mathrm{max}}/2$, and we have computed $r_{\mathrm{max}}$ up to 1000. 

Fig.~\ref{fig:axfig1} (a) describes the dispersion relation of the linear SWT along the momentum path $\Gamma\rightarrow X\rightarrow M\rightarrow \Gamma$ with $\alpha$ changing from $\alpha=2.0$ to $4.0$. For large $\alpha$, the LR coupling rapidly decays which makes its disperion relation of single magnon similar to  that of the typical anti-ferromagnetic square lattice only with nearest-neighbor coupling. Here, the single magnon dispersion relation is gapless only at $\Gamma$ and $M$. As $\alpha$ decreases, the LR couplings strongly distort the dispersion relation, which makes the magnon excitation cost more energy and the dispersion relation goes higher as $\alpha$ decreasing. But this dispersion relation obtained from the linear SWT theory still remains gapless at $\Gamma$ and $M$. Actually, the single magnon dispersion relation would become discrete at $\Gamma$ and $M$ in the limit $r_{\mathrm{max}}\rightarrow\infty$.

With $\Delta \mathbf{q}$ the relative momentum away from the $M$ point, we plot the dispersion relation along the $M\rightarrow\Gamma$ direction in Fig.~\ref{fig:axfig1} (b) with a double logarithmic scale at $\alpha=3.5$ with varying $r_{\mathrm{max}}$, which shows the power-law dependence between $\Delta \vec{q}$ and $\omega$. We fit the linear SWT results with $\omega_{\mathbf{q}}=A|\Delta \vec{q}|^{s}$ near the $M$ point. 
Note that for small $r_{\mathrm{max}}$ ($r_{\mathrm{max}}\leq100$), the single magnon dispersion around the $M$ point still depends on $r_{\mathrm{max}}$, which is shown in Fig.~\ref{fig:axfig1} (b) with $r_{\mathrm{max}}$ changing from $16$ to $1000$. Such a dependence would disappear and $s$ would finally converge in the limit $r_{\mathrm{max}}\rightarrow\infty$. 
In order to demonstrate this convergence process as $r_{\mathrm{max}}\rightarrow\infty$, we plot two black dashed lines that $\omega_{\vec{q}}\propto|\Delta\vec{q}|^{0.48}$ in (b), where $s=0.48$ comes from the fitting of the linear SWT result with $r_{\mathrm{max}}=1000$. 
In Fig.~\ref{fig:axfig1} (b), as $r_{\mathrm{max}}$ increases, $s$ converges to $0.48$. 
Finally, with $r_{\mathrm{max}}=1000$, Fig.~\ref{fig:axfig1} (c) presents our fitting about the relation between the power $s(\alpha)$ and the decay exponent $\alpha$, which suggests the limit $r_{\mathrm{max}}\rightarrow\infty$ and is also plotted as the blue dots in Fig.~\ref{fig:fig1} (d) in the main text.


Similarly, we also plot our linear SWT results of the ferromagnetic case in Fig.~\ref{fig:axfig1} (d), (e), and (f). Fig.~\ref{fig:axfig1}(f) also shows the relation between the power $s(\alpha)$ and $\alpha$ taking $r_{\mathrm{max}}=1000$, which is also given as the blue dots in Fig.~\ref{fig:fig1} (e) in the main text.

{\color{black}
	In addition, the Hamiltonian Eq.~\eqref{ferroH} and ~\eqref{antiferroH} are super-extensive for $\alpha\le2$ and their ground state energy diverges super-linearly with the number of sites $N$. Historically, to suppresses such a divergence of the ground state energy, the Kac construction has been proposed\cite{VanKac1963,MickaelClusting1995,DefenuMetastability2021,MichaeUniversal2013}. By coupling a normalization factor $\mathcal{N}$ to the Hamiltonian, Eq.~\eqref{ferroH} and ~\eqref{antiferroH} becomes
	\begin{eqnarray}
		H_{FM}^{Kac}&=&-\mathcal{N}J\sum_{i\neq j}\frac{1}{|\vec{r_{i}-\vec{r_{j}}|^{\alpha}}} \vec{S_{i}}\cdot
		\vec{S_{j}},
		\label{ferroHKac}\\
		H_{AFM}^{Kac}&=&\mathcal{N}J\sum_{i\neq j}\frac{(-1)^{|x_i+y_i-x_j-y_j+1|}}{|\vec{r_{i}-\vec{r_{j}}|^{\alpha}}} \vec{S_{i}}\cdot
		\vec{S_{j}},
		\label{antiferroHKac}\\
		\mathcal{N}&=&\frac{N-1}{\sum_{i\ne j}\frac{1}{\left | r_i-r_j \right | ^{\alpha}}}.
		\label{Kac}
	\end{eqnarray}
	For convenience of linear SWT analysis, we rewrite $\mathcal{N}$ by
	\begin{eqnarray}
		\mathcal{N}=\frac{N-1}{N}\frac{1}{\sum_{j}\frac{1}{2\left | r_{i_0}-r_j \right | ^{\alpha}}}.
		\label{Kacre}
	\end{eqnarray}
	Since $\lim\limits_{N\rightarrow\infty}\frac{N-1}{N}=1$, Eq.~\eqref{Kacre} actually introduces a factor $1/\left[\sum_{j}\frac{1}{2\left | r_{i_0}-r_j \right | ^{\alpha}}\right]$ to the Hamiltonian of the linear SWT in the momentum space (like Eq.~\eqref{appLSW:Ham}). 
	
	To see whether the normalization factor Eq.~\eqref{Kac} influences the excitation gap in the Higgs regime, we calculate the spin wave dispersion $\omega(\Delta \vec{q}=2\pi/r_{\mathrm{max}})$ at the momentum point $\Gamma+\Delta {\vec{q}}$ for the ferromagnetic case ($M+\Delta {\vec{q}}$ for the anti-ferromagnetic case) with $r_{\mathrm{max}}\rightarrow\infty$, which is plotted in Fig.~\ref{fig:axfig2}. Fig.~\ref{fig:axfig2}(a) stands for the staggered antiferromagnetic coupling, while (b) is for the ferromagnetic case. $\alpha$ ranges from $1.5$ to $1.7$ in both cases. As it is shown, for both of the cases, $\omega(2\pi/r_{\mathrm{max}})$ converges to a finite constant as $r_{\mathrm{max}}\rightarrow\infty$ even with the normalization factor from Kac construction. Such a behavior proves the spectra in the Higgs regime remain gapped even with the Kac construction, and our conclusion is not affected by the divergence of the ground state energy.
	
	\begin{figure}[htp]
		\includegraphics[width=0.85\columnwidth]{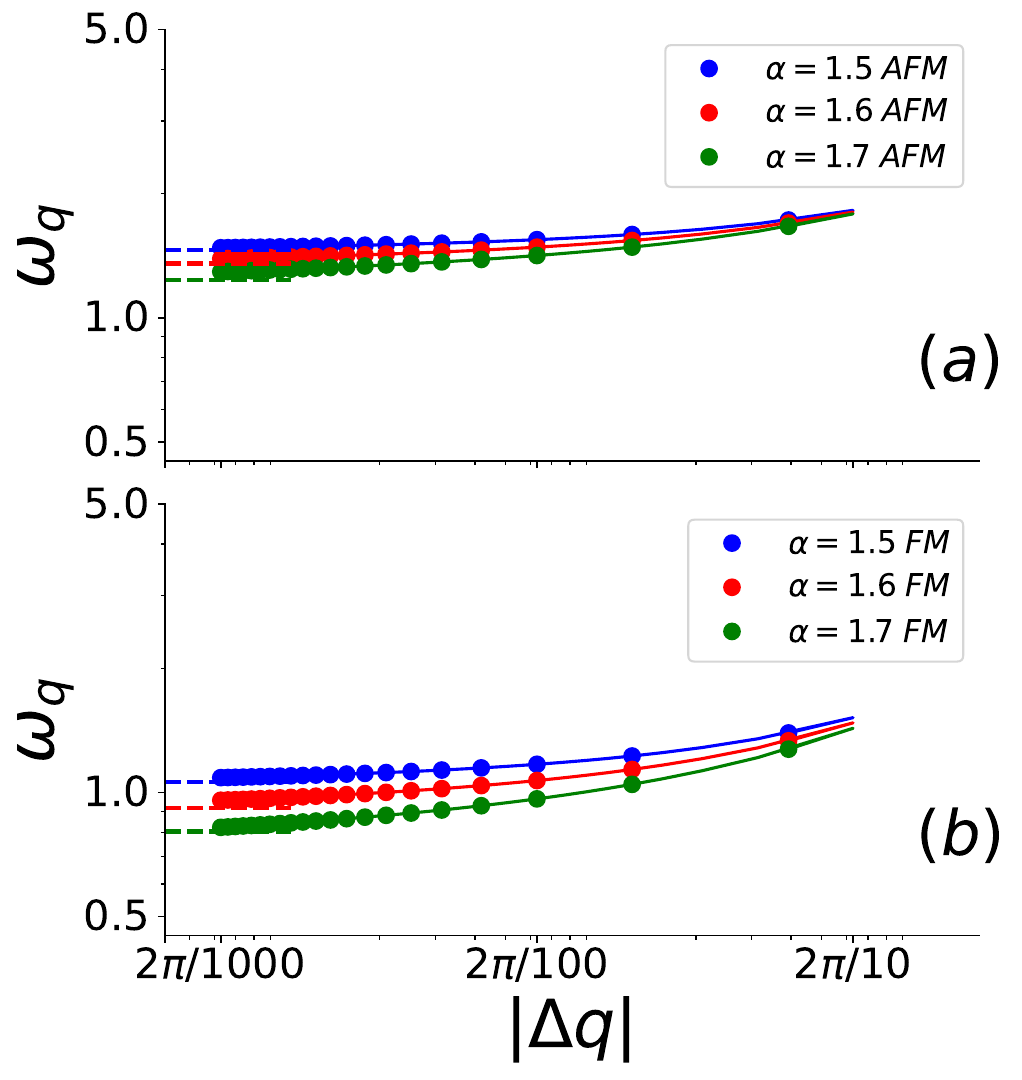}
		\caption{{\color{black}The linear SWT results considering the Kac normalization. Panel (a) is for the the staggered antiferromagnetic coupling and describes the spin wave dispersion relation near the $M$ point while (b) stands for the ferromagnetic case and near the $\Gamma$ point. $|\Delta\vec{q}|=2\pi/r_{\mathrm{max}}$ with $r_{\mathrm{max}}$ ranging from $16$ to $1000$. The blue, green and red line refer to $\alpha=1.5,\;1.6,\;1.7$. The dashed line is the energy gap $\Delta$ observed by fitting $\omega(\Delta\vec{q})$ with $a|\vec{q}|^b+\Delta$.}}
		\label{fig:axfig2}
	\end{figure}

\begin{figure}[htp]
	\includegraphics[width=\columnwidth]{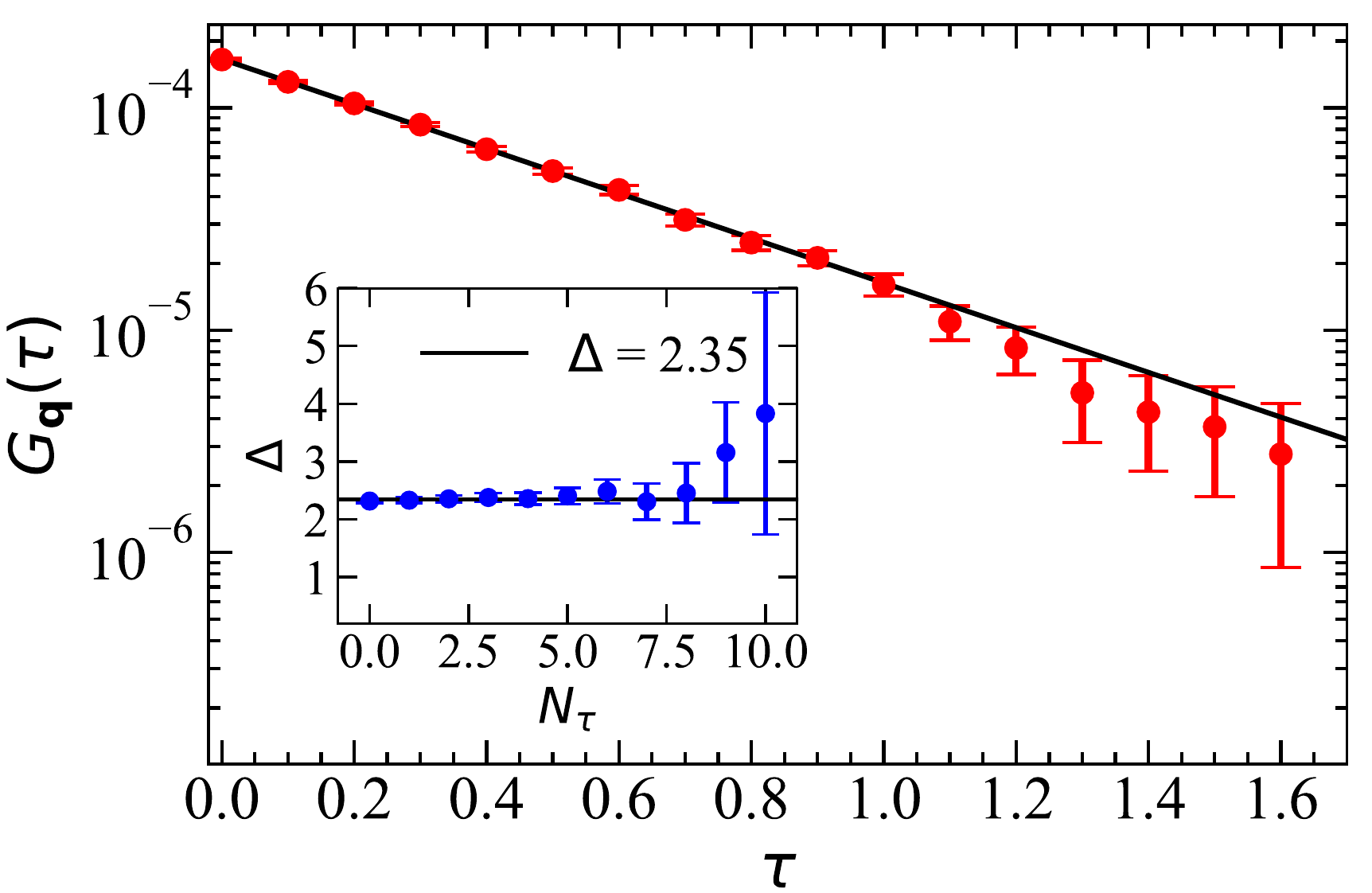}
	\caption{The fitting of energy gap with the data of the correlation function $G_{\vec{q}}(\tau)$ versus $\tau$ for $L=64$ and $\alpha=2.5$ at $\vec{q} = (3\times 2\pi/L,0)$ for the ferromagnetic case. The inset shows the obtained gap when the first $N_{\tau}$ data points are omitted before fitting.}
	\label{fig:sm1}
\end{figure}

\subsection{Fitting with QMC data}
In order to obtain the low-energy spectra of $H_{FM}$ and $H_{AFM}$ in QMC~\cite{SandvikQuantum1991,SandvikStochastic1999,yanEmergent2023,yanWormhole2021}, we compute the imaginary time correlation function $G_{\vec{q}}(\tau)\equiv \langle S^{z}_{\vec{q}}(\tau)S^{z}_{-\vec{q}}(0)\rangle-\langle S^{z}_{\vec{q}}\rangle^{2}$, where $ S^{z}_{\vec{q}}\equiv \frac{1}{\sqrt{N}}\sum_{\vec{r}}e^{i\vec{q}\cdot\vec{r}}S^{z}_{\vec{r}}$. Here we consider the periodic boundary conditions in the simulation so that $(q_x,q_y)=(\pm \frac{2\pi m}{L},\pm \frac{2\pi n}{L})$ with $m$ and $n$ being integers are physical momenta on a $L\times L$ square lattice. The loop update scheme of SSE QMC is purposely adapted to cope with the long-range interactions by assigning each bond with a separate bond weight and bond type (ferromagnetic or antiferromagnetic)~\footnote{The diagonal and off-diagonal operators sit on the same bond are still equal weighted even now, different weights and type bonds are assigned to different bonds, and this ensures that the loop update scheme does not need to be amended.}. Since
\begin{equation}
\begin{aligned}
\langle S^{z}_{\vec{q}}(\tau) S^{z}_{-\vec{q}}(0)\rangle & =\left\langle\mathrm{e}^{H \tau} S^{z}_{\vec{q}}(0) \mathrm{e}^{-H \tau} S^{z}_{-\vec{q}}(0)\right\rangle= \\
\left(\sum_{l=0} \mathrm{e}^{-\beta E_l}\right)^{-1} & \times \sum_{n, m=0}|\langle n|S^{z}_{\vec{q}}| m\rangle|^2 \mathrm{e}^{-\left(E_m-E_n\right) \tau} \mathrm{e}^{-\beta E_n}
\end{aligned}
\end{equation}
where $H|n\rangle=E_n|n\rangle$ and $E_0$ is the ground state energy of the system. When $ \beta\Delta E_1\gg 1$ where $\Delta E_n=E_n-E_0$, we can estimate
\begin{equation}
G_{\vec{q}}(\tau)\approx
\sum_{n=1}|\langle 0|S^{z}_{\vec{q}}| n\rangle|^2 \left(\mathrm{e}^{-\Delta E_n(\vec{q}) \tau}+\mathrm{e}^{-\Delta E_n(\vec{q})(\beta-\tau)}\right).
\end{equation}
When the imaginary time is sufficiently large, we assume that the system will gradually evolve to the ground state, so that the correlation function can be further approximated by
\begin{equation}
G_{\vec{q}}(\tau)\approx
|\langle 0|S^{z}_{\vec{q}}| 1\rangle|^2 \mathrm{e}^{-\Delta E_1(\vec{q})\tau}.
\end{equation}
If $|\langle 0|S^{z}_{\vec{q}}| 1\rangle|^2$ is finite (which is usually the case), we can then extract the energy gap for each $\vec{q}$ point by fitting $G_{\vec{q}}(\tau)$ with an exponentially decaying function.}

We fit the QMC data of $G_{\vec{q}}(\tau)$ by the relation $G_{\vec{q}}(\tau)\propto e^{-\Delta_{\vec{q}}\tau}$ and the fitting process is shown in Fig.~\ref{fig:sm1}. We first choose the data points for fitting according to their relative errors. If the relative error of one data point is less than 0.2, then the data point is chosen to be used for fitting. In the fitting process, we gradually omit the first $N_{\tau}$ data points and then do the curve fitting to find the most probable gap. As shown in the inset of Fig.~\ref{fig:sm1},  the fitting error becomes intolerant when $N_\tau=10$ and the fitted gap converges around $\Delta=2.35$ when $N_\tau$ gradually decreases to 0. In this case, we choose $\Delta=2.35$ to be the fitted gap for the data. Note that we find that for all the $\mathbf{q}$ points at different $\alpha$, the fitted gap does not change evidently with $N_\tau$, which means that higher excited states have much bigger energy gaps then the first excited states($\Delta E_{2}\gg\Delta E_{1}$) so that $e^{-\Delta E_{1}\tau}$ term in $G_{\vec{q}}(\tau)$ contributes much more than other terms for the range of $\tau$ we consider.

\begin{figure}[htp]
	\includegraphics[width=\columnwidth]{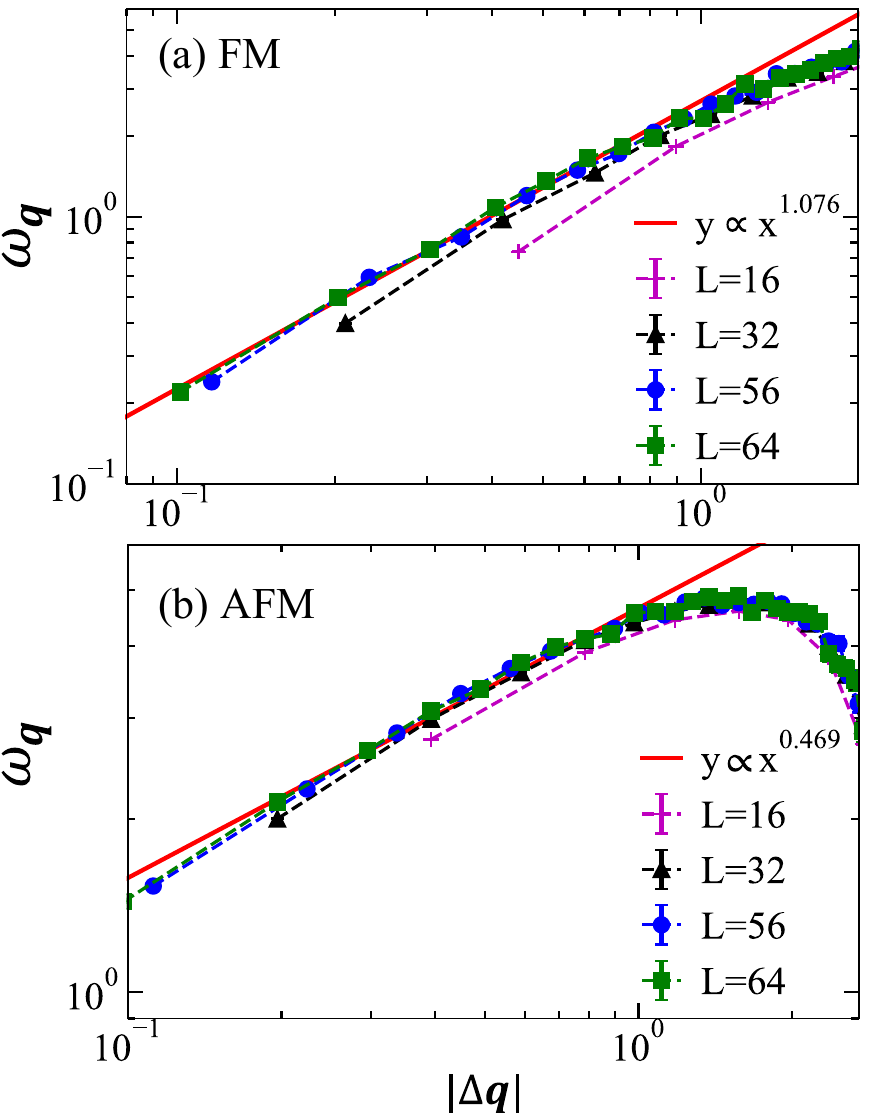}
	\caption{Dispersion relation at $\alpha=3$ for various system sizes $L$. (a) Dispersion of $H_{FM}$ near $\Gamma$ with $\Delta{\mathbf{q}}$ denotes the relative momentum away from $\Gamma$. Red line $y\propto x^{1.076}$ shows the fitted power $s(\alpha=3)=1.076$ using $L=64$ QMC data. (b) Dispersion of $H_{AFM}$ near $M$ with $\Delta{\mathbf{q}}$ denotes the relative momentum away from $M$. Red line $y\propto x^{0.469}$ shows the fitted power $s(\alpha=3)=0.469$ using $L=64$ QMC data.}
	\label{fig:sm2}
\end{figure}

Fig.~\ref{fig:sm2}(a) shows the dispersion of $H_{FM}$ near $\Gamma$ and Fig.~\ref{fig:sm2}(b) shows the dispersion of $H_{AFM}$ near $M$ for various system sizes $L$ at $\alpha=3$. $|\Delta\mathbf{q}|$ denotes the relative momentum away from the $\Gamma$ in (a) ($M$ in (b)). Plotting under double logarithm scale, it is demonstrated that the power of low-momentum dispersion $s(\alpha)$ depends on the system size $L$. However, as the system size $L$ increases, the dispersion gradually converge and $s(\alpha)$ will finally remain unchanged as $L\to \infty$. Such process can already be seen as the vast majority of $L=56$ and $L=64$ data collapse onto the same curve in both FM and AFM cases, we thus obtain $s(\alpha)$ by fitting $L=64$ data and $y\propto x^{s(\alpha)}$ is plotted as red lines in Fig.~\ref{fig:sm2}. The same red lines are shown in the insets (c), (d) and (e) in Figs.~\ref{fig:fig2} and ~\ref{fig:fig3} in the main text.

\end{document}